\def\p{\partial}
\documentstyle[12pt]{article}
\begin{document}

\begin{titlepage}
\nopagebreak
\begin{flushright}

hepth@xxx/9305152
\\
ETH-TH/93-21
\\

May  1993
\end{flushright}

\vglue 2.5  true cm
\begin{center}
{\large\bf
$W_{\infty}$--GEOMETRY \\ AND \\ ASSOCIATED CONTINUOUS
TODA SYSTEM\footnote{
\LaTeX \ file  available from
hepth@xxx.lanl.gov (\# 9305152)}}

\vglue 1  true cm
{\bf Mikhail V.~ SAVELIEV}\footnote{ On leave of absence from
the Institute for High Energy Physics,
142284, Protvino, Moscow region, Russia.\quad
E-mail: SAVELIEV@M10.IHEP.SU }\\
{\footnotesize Institute for Theoretical Physics, ETH--H\"onggerberg\\
CH--8093 Z\"urich, ~Switzerland.\\
and
} \\
{\bf Svetlana A.  SAVELIEVA}\\
{\footnotesize Institute for High Energy Physics,\\
142284, Protvino, Moscow region, ~Russia.}\\
\medskip
\end{center}

\vfill
\begin{abstract}
\baselineskip .4 true cm
\noindent
We discuss an infinite--dimensional k\"ahlerian manifold associated
with the area--preserving diffeomorphisms on two--dimensional torus,
and, correspondingly, with a continuous limit of the $A_r$--Toda
system. In particular, a continuous limit of the $A_r$--Grassmannians
and a related Pl\"ucker type formula are introduced as relevant notions for
$W_{\infty}$--geometry of the self--dual Einstein space with the
rotational Killing vector.
\end{abstract}
\vfill
\end{titlepage}
\baselineskip .5 true cm

1. A remarkable correspondence between $W$--geometry of the two--dimensional
$A_r$--Toda system and the Pl\"ucker embedding has been
recently discovered in \cite{GM}, where it is shown that
the K\"ahler potentials of the pseudo--metrics induced on the corresponding
$W$--surfaces coincide with the $A_r$--Toda
fields.\footnote{Here by a $W$--surface the authors mean a
two--dimensional manifold  with a chiral embedding into ${\bf
C}^N$ with an emphasis on its extrinsic geometry.}Note that this
fact takes place \cite{GS} also for the $W$--surfaces associated with an
arbitrary finite--dimensional simple Lie algebra ${\cal G}$ endowed with the
canonical gradation, and the corresponding ${\cal G}$--Toda
fields. In turn this relation gives an independent proof of
the infinitesimal Pl\"ucker type formula for the curvatures of the
pseudo--metrics in terms of the Cartan matrix $k$ of
${\cal G}\equiv {\cal G}(k)$.
\footnote{The statement for an arbitrary finite--dimensional simple Lie
algebra ${\cal G}(k)$, which generalizes the well--known
Pl\"ucker formula for the $A_r$--Grassmannians, see e.g., \cite{GH},
have been conjectured in \cite{G} and proved in \cite{P} using the
algebraic geometry methods; see also \cite{Y}.}

It seems reasonable to study such a relation for $W_{\infty}$--geometry,
i.e., in a continuous limit \cite{S1},\cite{KSSV},\cite{S2}
\begin{equation}
\p ^2x(z_+,z_-;\tau )/\p z_+\p z_-=\exp \p ^2x(z_+,z_-;
\tau )/\p \tau ^2,
\label{f1}
\end{equation}
of the
$A_r$--Toda system  describing, in particular, the self--dual Einstein
space with the rotational Killing vector \cite{BF}, and associated  with the
area--preserving diffeomorphisms $S_0\mbox{ Diff }T^2$ on two--dimensional
torus $T^2$; and even for more general dynamical systems like those
associated with a continuum Lie algebra with the Cartan operator
${\cal K}$ \cite{SV}.\footnote{The axiomatics of the continuum
Lie algebras naturally follows the brilliant formulation of the
contragredient Lie algebras of finite growth (Kac--Moody algebras)
given by V. Kac \cite{K1}, while {\it continuum Lie algebra possesses,
in general, an infinite--dimensional Cartan subalgebra and a contiguous
set of roots}, see \cite{SV}.}

A limiting continuous procedure \`a la Volterra method proved itself
in a good light as a hint in constructing the continuum Lie algebras.
Now, natural step is to study, using similar reasonings,
the infinite--dimensional k\"ahlerian manifold and $W_{\infty}$--geometry
associated with system (\ref{f1}), and endowed with the structure of the
$A_{\infty}$ algebra. Note, that there is a number of papers, see e.g.,
\cite{K2} and references therein, where
infinite--dimensional K\"ahler geometry and Grassmannians
associated with the group of smooth based loops on a connected
compact Lie group were investigated. However, there the arising (flag)
manifolds behave in many respects like finite--dimensional ones,
while those under consideration in the present paper deal with
the notions and objects of a novel nature which, as far as we know,
have not been treated before. To be honest, let us mention from the very
begining that in our paper the word ``manifold '' for the continuous case
should be understood in an conventional sense, since our consideration here
is rather formal yet.

To proceed with our program, first
give briefly some information about a continuum formulation
of the algebras in question.

2. The algebra $A_{\infty}$ is isomorphic to
the algebra $S_0\mbox{ Diff }T^2$ which in turn, as the continuum Lie
algebra ${\cal G}(E; -\frac{d^2}{d\tau ^2};\mbox{ id })$ with the Cartan
operator $-d^2/d\tau ^2$, is isomorphic, in
accordance with \cite {SV}, to the Poisson bracket Lie algebra
on $T^2$. The last one, considered as {\bf Z}--graded continuum algebra
${\cal G}(E;-i\frac{d}{d\tau};-i\frac{d}{d\tau}) =
\mathop{\oplus}\limits_{m \in {\bf Z}}{\cal G}_m$ is defined by the
commutation relations
\begin{equation}
[X_m(\varphi),X_n(\psi)] = i X_{m+n}(n\varphi'{\psi}-m\varphi\psi').
\label{f2}
\end{equation}
Here $X_m(\varphi)=\int d\tau X_m (\tau )\,\phi (\tau )$ are the elements
of the subspaces ${\cal G}_m$
parametrized by the functions $\varphi(\tau)$ belonging to the
algebra $E$ of trigonometrical polynomials on a circle;
$\varphi' \equiv \frac{d\varphi}{d\tau}$. This algebra is of constant
(in functional sense) growth since
${\cal G}_n \simeq {\cal G}_1 \simeq E$; its Cartan subalgebra
$\wp \simeq {\cal G}_0$ is infinite--dimensional, the roots are
$n\delta'(\tau)$. Let $\wp^*$ be an algebra
dual to $\wp$, let $V$ be a ${\cal G}$-module and $\lambda \in \wp^*$.
Denote by $V_\lambda$ a set of vectors $v \in V$ satisfying
$X_0(\varphi)v = \lambda(\varphi)v$ for all $\varphi \in E$.
Moreover, it can be shown by an appropriate limit procedure
(starting from $A_r$ and using the aforementioned isomorphism
${\cal G}(E;-i \frac{d}{d\tau};-i \frac{d}{d\tau}) \simeq
{\cal G}(E;-\frac{d^2}{d\tau^2};\mbox{ id }) \simeq A_\infty$) that there
exists a nonzero vector ${\bar v} \in V$ such that
${\cal G}_{m}(\bar v) = 0$ for $m>0$ and $U({\cal G})(\bar v) = V$.
Here $U(\cal G)$ is the universal enveloping algebra for ${\cal G}$.
By analogy with the usual (``discrete'') case this ${\cal G}$-module
$V$ is called the highest weight module, and $\bar v$ the highest weight
vector. A symmetrical bilinear invariant form on the local part ${\cal
G}_{-1}\oplus {\cal G}_0\oplus {\cal G}_{+1}$ of the algebras in
question is defined as follows
\[
\mbox{ tr }(X_i(f)X_j(g)) = \delta _{i+j,0}(f, g), \qquad i,j=0, \pm 1; \]
where
\begin{displaymath}
{(f, g)} = \int d\tau \begin{array}{c}
f(\tau )g(\tau )\\ \\
\frac{df(\tau )}{d\tau }\,\frac{dg(\tau )}{d\tau }
\end{array} \mbox{ for }
\begin{array}{c}
g(E; -i\frac{d}{d\tau}; -i\frac{d}{d\tau})\\ \\
g(E; -\frac{d^2}{d\tau^2};\mbox{ id }).
\end{array}
\end{displaymath}

In what follows a continuous version (for $A_{\infty}$) of the
highest weight vectors of the fundamental representations of $A_r$
is denoted by $\vert \tau >$, for which
\begin{eqnarray}
X_0(\phi )\,\vert \tau >=\phi (\tau )\vert \tau >;& & \;
X_m(\phi )\,\vert \tau >=0 \mbox{ for } m>1; \nonumber \\
\mbox{ and } & &
X_{-1}(\tilde{\tau })\vert \tau >=0 \mbox{ for } \tilde{\tau }
\not = \tau .
\label{f3}
\end{eqnarray}

3. A relevant object for the description of the
$A_r$--$W$ - geometry of ${\bf C}^r$ - target manifolds with a positive
curvature form is the Pl\"ucker embeddings of the Grassmannians ${\cal
G}r(N\vert \, p)$ in the projective spaces,
${\cal G}r(N\vert \, p)\Rightarrow {\bf CP}^{{ N \choose p }\,
-1}$. Here inhomogeneous coordinates $Z_{ij}=\mbox{ det }{\cal R}_{a-i+j}
/\mbox{ det }{\cal R}_{a}$ of this map are the determinants
$\mbox{ det }{\cal R}_a$ of all the $p\times p$ minors ${\cal R}_a$ of
the element ${\cal R}$ of ${\cal G}r(N\vert \, p)$ in its matrix realization.
A natural continuous version of the Pl\"ucker representation for the
pseudo--metric in terms of the norms $\vert {\cal R}_{p}\vert
=\exp (-{\cal Q}_p)$, and the curvature form $R_p$ of the
pseudo--metric $ds^2_p$ (infinitesimal Pl\"ucker formula),
\begin{equation}
ds^2_p  =  \frac{i}{2} \,\partial \bar {\partial }\,\mbox{ log }
\vert {\cal R}_{p}\vert ^2
= \frac{i}{2}\,\prod \limits _q e^{2k_{pq}{\cal Q}_q}\,
dZ\wedge d\bar {Z};\quad R_p \,=\,\sum _q k_{pq}ds^2_q,
\label{f4}
\end{equation}
can be written as
\begin{equation}
ds^2 (\tau )  = \frac{i}{2}\,\p \bar {\p }{\cal Q}(\tau )=
\frac{i}{2}\, e^{2\p ^2{\cal Q}(\tau )/\p \tau ^2}\,
dZ\wedge d\bar {Z};\quad R (\tau ) \,=\,\frac{\p ^2}{\p \tau ^2}
\,ds^2(\tau ).
\label{f5}
\end{equation}
Of course, at this stage our arguments are very formal since
the objects which we have introduced are not yet specified enough,
and we need to provide some parametrization for them to clarify
their group--algebraic meaning.

For this goal let us recall that for $A_r$--$W$ - geometry \cite{GM}, the
trivial ${\bf C}^N$ - target manifolds are simply ${\bf C}^N$ Riemannian
manifolds with $2N$ real dimensions and with the homogeneous (Euclidean)
coordinates $Y^A$ and $\bar{Y}^{\bar {A}}$, $0\leq A, \bar {A} \leq N-1$,
such that the linear element  $d\,s^2=\sum _A dY^A\,d\bar {Y}^A$. The
corresponding ${\bf C}^N$-- $W$ - surface is a two-dimensional manifold with
a chiral embedding into ${\bf C}^N$, which is defined by the independent
functions $Y^A(z_-)$ and $\bar {Y}^{\bar {A}}(z_+)$, $0\leq A, \bar {A}
\leq N-1$,\footnote{In other words, one does not assume here that
$\bar {Y}^{\bar {A}}$ is the complex conjugate of $Y^A$.}
with $N$ being the dimension of the corresponding representation
of $A_r$. Here an explicit parametrization of the cosets that are
k\"ahlerian manifolds associated with the fundamental
representations of $A_r$ with the highest weight states $\vert
\lambda _i>,\,1\leq i\leq r$, is based on the following scheme.
Denote by $S_i\subset A_r$ the stability group of $\vert \lambda _i>$.
Then the Grassmannian ${\cal G}r(N\vert \, i)\equiv A_r/S_i$
is parametrized by the formula
$\exp (\sum  Y^A\,{\cal F}_A)
\vert \lambda _i>;\, 1\leq A\leq \mbox{ dim }A_r/{\cal S}_i\;
{\cal F}_A \in A_r/{\cal S}_i$,
and the exponential of the corresponding K\"ahler potential is
the highest weight matrix element
\[
\exp {\cal K}_i=<\lambda _i\vert \exp (-\sum \limits _{A}\bar {Y}^A{\cal
F}_A^{\dagger})\,\exp (\sum \limits _{A}Y^A{\cal F}_A)\vert
\lambda _i>. \]
On the  $W$--surface associated with the  dynamical system, here -- the
Abelian $A_r$--Toda system, this matrix element (realizing the highest
vector of the $i$-th fundamental representation of $A_r$) is given by
the corresponding $\tau $--function, and $-{\cal K}_i  =  {\cal Q}_i$
coincide \cite{GM} with the $A_r$--Toda fields, whose general solution
is determined, in accordance with \cite{LS1}, by $2r$ arbitrary
functions of $z_+$ and $z_-$ entering the corresponding tau--function.
\footnote{For the general solution to ${\cal G}$--Toda system associated
with an arbitrary finite--dimensional simple Lie algebra, see \cite{LS1}
and also \cite{LS2}.}

A realization of the object which seems to be natural to call a continuous
version of the Grassmannian, ${\cal G}r(E\vert \, \tau )$, deals with the
homogeneous coordinates $Y(z_-,\tau )$ and $\bar {Y}(z_-,\tau )$ of the
manifold which arise as continuous analogues of the euclidean coordinates
of the finite--dimensional picture. With account of definition
(\ref{f3}) of the highest weight state $\vert \tau >$, the exponential of
the K\"ahler potential ${\cal K}(\tau )$ for the infinite--dimensional
manifold under consideration is related to the continuous Toda
field $x(z_+, z_-;\tau )$ satisfying equation (\ref{f1}), by the
formula
\begin{eqnarray}
\exp {\cal K}(\tau ) & = & \vert {\cal R}(\tau )\vert =
e^{-x(z_+,z_-;\tau )} \nonumber \\
& = & e^{K^{-1}\mbox{ log }[\phi _-(z_-,\tau )
\phi _+(z_+,\tau )]}\,<\tau \vert {\cal M}_+^{-1}{\cal M}_-\vert
\tau >,
\label{f6}
\end{eqnarray}
which is a direct continualization of the corresponding $A_r$--Toda system.
Here $K\equiv -\frac{d^2}{d\tau ^2}$, and the dependence on
$z_{\pm}$ is omitted for
brevity. The functions ${\cal M}_{\pm}(z_{\pm})$ are defined as the solution
(multiplicative integral) of the initial-value problem for the equations
\begin{equation}
\p _{\pm}{\cal M}_{\pm}={\cal M}_{\pm}\,X_{\pm 1}(\phi _{\pm}).
\label{f7}
\end{equation}
The matrix element $<\tau \vert {\cal M}_+^{-1}{\cal M}_-\vert \tau >$
realizes the continuous version of the tau--function
depending on two arbitrary functions $\phi _{\pm}(z_{\pm},\tau )$
which determine the general solution \cite{S1},\cite{KSSV},\cite{S2} to
equation (\ref{f1}).\footnote{For the solution of the Cauchy (initial value)
problem to equation (\ref{f1}), see \cite{SS}.}  \footnote{A different
representation of the
tau--function for equation (\ref{f1}) has been discussed in \cite{T}. For
a more general system of such a type, associated with the universal Whitham
hierarchy, the solutions determining by an infinite number of arbitrary
functions of two variables, $\tau $ and, say $z_++z_-$, are given
 in \cite{Kr}.} In terms of the solution to equation (\ref{f7}),
symbolically expressed by ${\cal Z}_{\pm}$--ordered exponential,
${\cal M}_{\pm}={\cal Z}_{\pm}\exp \int dz_{\pm}'\,X_{\pm 1}(\phi _{\pm}
(z_{\pm}')),\,z_{\pm}'\in [c_{\pm},
z_{\pm}],\,c_{\pm}=\mbox{ const }$,
the continuous version of the Grassmannian ${\cal G}r(E\vert \,
\tau )$ on the $W_{\infty }$--surface is written as
\begin{equation}
{\cal Z}_-\exp [\int \limits _{c_-}\limits
^{z_-}dz_{-}'\,X_{-1}(e^{K\Phi (z_{-}')})]\,\exp [-X_0(\Phi )]\,\vert
\tau >,\quad \Phi \equiv K^{-1}\mbox{ log }\phi _-.
\label{f8}
\end{equation}

Finally, in these terms a continuous version ${\cal I}(\tau )$ of a
topological characteristic  of the $A_r$-- $W$--surface, that is used
to be the instanton number in \cite{GM}, can be written as
\begin{equation}
{\cal I}(\tau )=\frac{i}{2\pi }\int dz_+dz_-\exp
\p ^2x(z_+,z_-;\tau )/\p \tau ^2
\label{f9}
\end{equation}
under appropriate conditions imposed on the functions
$\phi _{\pm}(z_{\pm})$ determining the general solution (\ref{f6}) for
system (\ref{f1}).

As we have already mentioned, our construction is quite formal and
requires to be comprehended in more detail. The main, but surmountable
difficulty here is caused by the fact that the exponential mapping of the
algebra ${\cal G}(E; -\frac{d^2}{d\tau^2}; \mbox{ id })\simeq A_\infty$
gives a differentiable structure which is weaker than that of a Lie group
in the classical sense. Some of these problems were considered and reviewed
in the paper of M. Adams, T. Ratiu, and R. Schmid in \cite{K2}, see also
references therein.

Moreover, it will be very
interesting to extend the consideration to the case of infinite--dimensional
manifolds associated with some other continuum Lie algebras with
invertible Cartan operators so to have parametrizations
analogous to (\ref{f6}) and (\ref{f8}).

\bigskip

Finishing up the paper, one of the authors (M.V.S) would like to thank
F. E. Burstall, A. T. Fomenko, J.-L. Gervais, I. M. Krichever, J. Moser and
A. V. Razumov for fruitful discussions.
He also wishes to acknowledge  the Institute of Theoretical
Physics, especially  A. Wipf, for kind hospitality
during his visit at ETH in Zurich where the present paper was completed.

\end{document}